# A Schrödinger potential involving $x^{2/3}$ and centrifugal-barrier terms conditionally integrable in terms of the confluent hypergeometric functions

**V.A. Manukyan[1], T.A. Ishkhanyan[1,2], and A.M. Ishkhanyan[1,2]**

[1] Russian-Armenian University, Yerevan, 0051 Armenia
[2] Institute for Physical Research, Ashtarak, 0203 Armenia

E-mail: tishkhanyan@gmail.com

The solution of the one-dimensional Schrödinger equation for a potential involving an attractive $\sim x^{2/3}$ and a repulsive centrifugal-barrier $\sim x^{-2}$ terms is presented in terms of the non-integer-order Hermite functions. The potential belongs to one of the five bi-confluent Heun families. This is a conditionally integrable potential in that the strength of the centrifugal-barrier term is fixed to $91\hbar^2/(72m)$. The general solution of the problem is composed using fundamental solutions each of which presents an irreducible linear combination of two Hermite functions of a scaled and shifted argument. The potential presents an infinitely extended confining well defined on the positive semi-axis and sustains infinitely many bound states.

## 1. Introduction

The exact analytic solutions in terms of special functions provided a useful tool in establishing the wave mechanics and further in exploring the basic paradigms of the new quantum theory [1-10]. The most known of such functions are the functions of the hypergeometric class, among which the Gauss ordinary $_2F_1$ and Kummer confluent $_1F_1$ functions are the two key ones. The most of the commonly used functions of mathematical physics, such as the Legendre, Lagrange, Hermite, Jacobi, Bessel, Whittaker functions, etc., are particular specifications, modifications or special limiting cases of these two.

The list of the Schrödinger potentials for which the problem is solved using hypergeometric functions is conventionally divided into two major classes. The first set of potentials, which are referred to as *exactly* solvable, involves the ones for which all potential parameters are varied independently. The classical harmonic oscillator and Coulomb potentials discussed by Schrödinger [1,2], and the three other independent classical hypergeometric potentials by Morse [3], Eckart [4] and Pöschl-Teller [5] as well as many particular cases of these potentials [6-10] are well known examples of this type. The list of independent exactly solvable potentials has recently been notably extended to include three more potentials: the third ordinary hypergeometric [11], and the inverse square root [12] and Lambert-W [13,14] confluent hypergeometric potentials.

The second set of hypergeometric potentials involves the ones for which some restrictions are imposed on the potential parameters (e.g., a parameter is fixed to a constant). These potentials are referred to as *conditionally* integrable. Among well known examples of this type, much discussed in the past, are the two confluent hypergeometric potentials by Stillinger [15] and the ordinary hypergeometric potential by Dutt et al. [16] (several other examples are presented in [17-20]).

The classical exactly solvable hypergeometric potentials [1-5] and their conditionally integrable extensions [15-20] have mainly been derived through an ansatz presenting a product of an elementary function and a hypergeometric function. It has recently been shown, however, that the currently obtained set of potentials almost exhausts the list of all possible cases for which the Schrödinger equation can be solved by this one-term ansatz [21]. A notable progress is then achieved if extending the ansatz to involve two or more hypergeometric functions. It has been understood that this extension suggests a much more flexible tool to derive new hypergeometric potentials. Indeed, it is this way that has led to the above-listed new exactly solvable potentials [11-14] as well as many new conditionally integrable potentials proposed during the past few years (see, e.g., [12,13,22-26]). A basic argument supporting the irreducible multi-term solutions involving two or more special functions comes up if one considers piecewise potentials for which each piece is solvable through the single-term ansatz [27-29]. Indeed, in order to match the solutions for adjacent intervals one needs to implicate the general solutions of the differential equations which for the second order equations (as the Schrödinger equation is) are linear combinations of two independent fundamental solutions. It is then immediately understood that even though each fundamental solution itself involves just one special function, the resulting solution will be of multi-term structure with more than one special function involved [29].

In this paper we present a new conditionally integrable potential of this type. This is a sub-potential of a six-parameter bi-confluent Heun potential by Lemieux and Bose [30] (see also [21]). The potential involves an attractive $V_1 x^{2/3}$ term with arbitrary strength $V_1$ and a fixed-strength repulsive centrifugal-barrier term $V_{cf}/x^2$, $V_{cf} = 91\hbar^2/(72m)$. The potential is defined on the positive semi-axis and sustains infinitely many bound states. The general solution of the problem for this potential is composed of two fundamental solutions each of which is given as an irreducible combination of two non-integer-order Hermite functions or, alternatively, of two Kummer confluent hypergeometric functions. Using the explicit solution, we deduce the exact equation for the discrete spectrum and present an accurate approximation for the energy eigenvalues.

## 2. The potential and the general solution

The 1D stationary Schrödinger equation for the wave-function $\psi(x)$ of a quantum particle in the field described by a potential $V(x)$ is written as

$$\frac{d^2\psi}{dx^2} + \frac{2m}{\hbar^2}(E - V(x))\psi = 0, \qquad (1)$$

where $m$ is the mass and $E$ is the energy of the particle, and $\hbar$ is Planck's constant. The conditionally integrable confluent hypergeometric potential we introduce is

$$V = \frac{91\hbar^2}{72mx^2} + V_0 + V_1 x^{2/3}. \qquad (2)$$

The potential belongs to the bi-confluent Heun family for which $m_1 = -1/2$ (see [21,25]). This family possesses several other known conditionally integrable confluent hypergeometric sub-potentials., among which the first and the most known one is the first Stillinger potential [15], Eq. (2.17). Other examples include the second Exton potential [31], Eq. (22), and the three-term potential that we have introduced very recently [25].

For positive $V_1$, the potential presents an infinitely extended well shown in Fig. 1. The potential can be applied to model the sub-nuclear interactions in particle physics by a confining fraction-power central potential $r^{2/3}$ [32]. It can also be applied in atomic, molecular and optical physics to describe the laser excitation of a two-state quantum system by a corresponding field configuration.

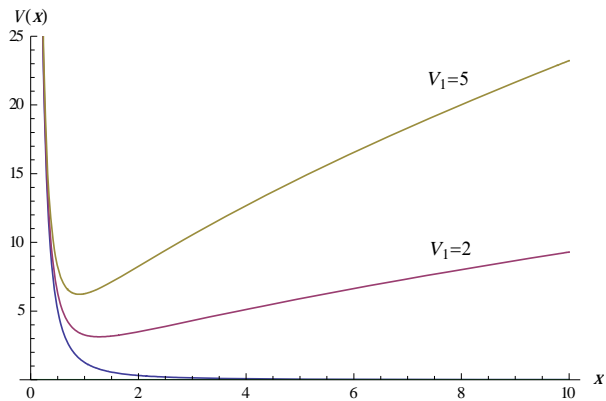

Fig.1 Potential (2) for positive $V_1$ ($m = \hbar = 1$, $V_0 = 0$). The lower curve corresponds to $V_1 = 0$

The bi-confluent Heun families of potentials are derived by reducing the Schrödinger equation to the bi-confluent Heun equation via the variable transformation given by the equation $dz/dx = z^{m_1}$ [33-35]. The families are identified by the exponent $m_1$, which adopts only five half-integer values: $m_1 = -1, -1/2, 0, +1/2, +1$. To derive the hypergeometric sub-potentials of these families, the solution is expanded in terms of the Hermite functions and the resulting series is terminated [25,36]. The termination on the fifth term for the family with $m_1 = -1/2$, if the potential parameters are allowed to be dependent, leads to the potential we introduce here. The solution of the Schrödinger equation (1) then presents a linear combination of five contiguous non-integer-order Hermite functions. This combination is further readily reduced to an irreducible combination of just two Hermite functions.

Though straightforward and merely technical, the calculation is slightly cumbersome. However, the result, which can be verified by direct substitution, is rather compact. It is as follows. If $V_1 > 0$ and $E > V_0$, a fundamental solution of the Schrödinger equation (1) for potential (2) is written as

$$\psi_F(x) = x^{-7/6} e^{-\left(\sqrt{2a}-z\right)^2/2} u(x), \tag{3}$$

where

$$u(x) = \left(\sqrt{2a}(1-2z^2) - z(3-2z^2)\right) H_{a+1/2}\left(\sqrt{2a}-z\right) - \left(1-z^2\right) H_{a+3/2}\left(\sqrt{2a}-z\right). \tag{4}$$

Here $z = \sqrt{3\varepsilon}\, x^{2/3}$ and $H_a$ is the Hermite function, and the involved parameters are given as

$$\varepsilon = \sqrt{\frac{mV_1}{2\hbar^2}}, \quad a = \frac{3m^2(E-V_0)^2}{32\hbar^4 \varepsilon^3}. \tag{5}$$

The second independent solution is constructed by changing $\varepsilon \to -\varepsilon$, $a \to -a$ in equations (3),(4). Hence, the general solution of the problem is written as a linear combination, with arbitrary constant coefficients $C_{1,2}$, of the two fundamental solutions:

$$\psi(x) = C_1 \psi_F + C_2 \psi_F \big|_{\varepsilon \to -\varepsilon, a \to -a}. \tag{6}$$

Using the representation [35]

$$H_\nu(z) = \sqrt{\pi}\, 2^\nu \left( \frac{{}_1F_1\left(-\frac{\nu}{2}; \frac{1}{2}; z^2\right)}{\Gamma\left(\frac{1-\nu}{2}\right)} - \frac{2z\, {}_1F_1\left(\frac{1-\nu}{2}; \frac{3}{2}; z^2\right)}{\Gamma\left(-\frac{\nu}{2}\right)} \right), \tag{7}$$

the solution can alternatively be written in terms of the Kummer confluent functions.

## 3. Bound states

As it was already mentioned above, for positive $V_1 > 0$ the potential presents an infinite confining well, hence, it sustains infinitely many bound states. We note that the polynomial reductions of solution (6), which are achieved for half-integer $a$, do not produce physically acceptable bound sates because they do not vanish in the origin. By requiring the wave function to vanish in the origin we obtain a relation for coefficients $C_{1,2}$:

$$C_2 = -e^{-2a} \frac{\sqrt{2a}\, H_{a+1/2}\left(\sqrt{2a}\right) - H_{a+3/2}\left(\sqrt{2a}\right)}{\sqrt{-2a}\, H_{-a+1/2}\left(\sqrt{-2a}\right) - H_{-a+3/2}\left(\sqrt{-2a}\right)} C_1. \tag{8}$$

The equation for bound-state energy levels is then obtained by requiring the wave function to vanish at the infinity. After some simplifications, this yields the exact equation

$$\sqrt{2a}\, H_{a+1/2}\left(-\sqrt{2a}\right) + H_{a+3/2}\left(-\sqrt{2a}\right) = 0. \tag{9}$$

By its structure, this equation has much in common with that for the inverse-square-root potential [12]. Therefore, following the lines of the treatment there, we apply the identity $H_{\nu+1}(z) = 2zH_\nu(z) - 2\nu H_{\nu-1}(z)$ to rewrite the spectrum equation in the following equivalent form:

$$(1+2a) H_{a-1/2}\left(-\sqrt{2a}\right) + \sqrt{2a}\, H_{a+1/2}\left(-\sqrt{2a}\right) = 0. \tag{10}$$

Now, we observe that the involved Hermite functions have such indexes and arguments that are close to the *left* transient point $z \approx -\sqrt{2\nu}$ [37]. An appropriate approximation suitable for this region was suggested in [38]. Using that, we obtain the following approximation:

$$F \equiv 1 + \frac{\sqrt{2a}\, H_{a+1/2}\left(-\sqrt{2a}\right)}{(1+2a) H_{a-1/2}\left(-\sqrt{2a}\right)} \approx \frac{a^{2/3}}{3(1+2a) B_0} \left( \frac{3B_0}{a^{2/3}} - \frac{\sin(\pi a - \pi/3)}{\sin(\pi a + \pi/3)} \right), \tag{11}$$

where

$$B_0 = \frac{\Gamma(1/3)}{6\sqrt[3]{3}\,\Gamma(2/3)} \approx \frac{1}{5}. \tag{12}$$

This is a rather accurate result that well models the auxiliary function $F(a)$ in the whole permissible variation range of the parameter $a$ (see Fig. 2, where the exact numerical values are shown by filled circles).

It then follows from this result that the spectrum equation $F = 0$ is with high accuracy approximated by the transcendental equation

$$\frac{3/5}{a^{2/3}} - \frac{\sin(\pi a - \pi/3)}{\sin(\pi a + \pi/3)} = 0. \tag{13}$$

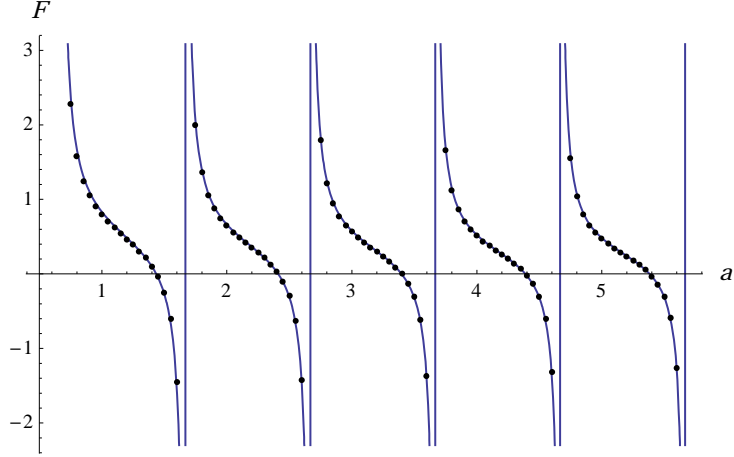

Fig. 2. Solid curves - approximation (11), filled circles - numerical result.

Even without appealing to analytical calculations, it is understood that the first term of this equation is small. To see this, we note that the roots of the spectrum equation (10) are close to positive half-integers (see Fig.2). The smallest root $a_0 = 1/2$ of the exact equation (9) (not shown in the Fig.2) does not produce a bound state because it leads to an identically zero wave function $\psi(x) \equiv 0$. The first root that works is $\approx 3/2$. The corresponding bound state is given by a wave function having one extremum, hence, this is the ground state to which we prescribe the number $n = 1$. Thus, $a_n \geq 1.5$ and the first term of equation (13) is small enough to be neglected in the first approximation. We then obtain $a_n \approx n + 1/3$, which provides, through the second equation (5), the semiclassical limit for large $n \to \infty$: $E_n \sim (n + 1/3)^{1/2}$, the Maslov correction index being 1/3 [38,39]. Indeed, for potential (2) the effective angular momentum such that $l(l+1) = 91/36$ is $l = 7/6$, so that the semiclassical result is $\gamma = (2l - 1)/4 = 1/3$. The relative error of this approximation is shown in Fig.3.

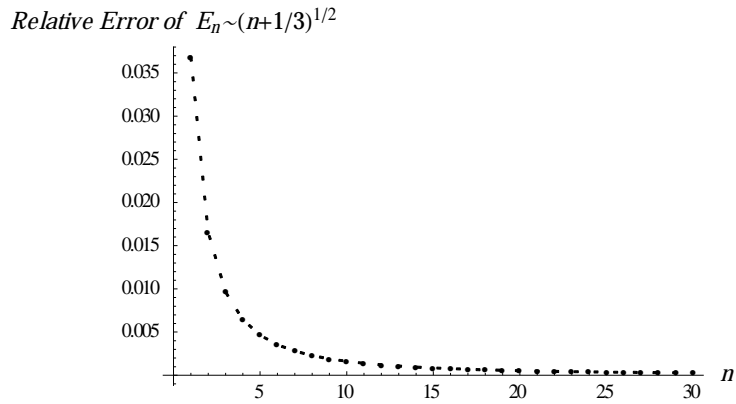

Fig. 3. Relative error of the semiclassical result $E_n \sim (n + 1/3)^{1/2}$.

Replacing the first term by $v = (3/5)(n+1/3)^{-2/3} \ll 1$, equation (13) is readily treated by the series expansion $a_n = n+1/3+b_1 v + b_2 v^2 +\ldots$. The result with the first two expansion terms straightforwardly leads, through the second equation (5), to the approximate spectrum

$$E_n \approx \left(\frac{128\hbar^2 V_1^3}{9m}\right)^{1/4} \left(n+\frac{1}{3}+\frac{1/6}{(n+1/3)^{2/3}}-\frac{1/20}{n+1/3^{4/3}}\right)^{1/2}, \quad n=1,2,3,\ldots \quad (14)$$

This is a rather advanced approximation. For the ground state, the relative error is $\approx 1.5 \times 10^{-4}$ (Fig. 4). Compared with the semiclassical result, for the first levels we have two orders of magnitude improvement in the accuracy. Fig. 5 depicts the first three normalized wave functions.

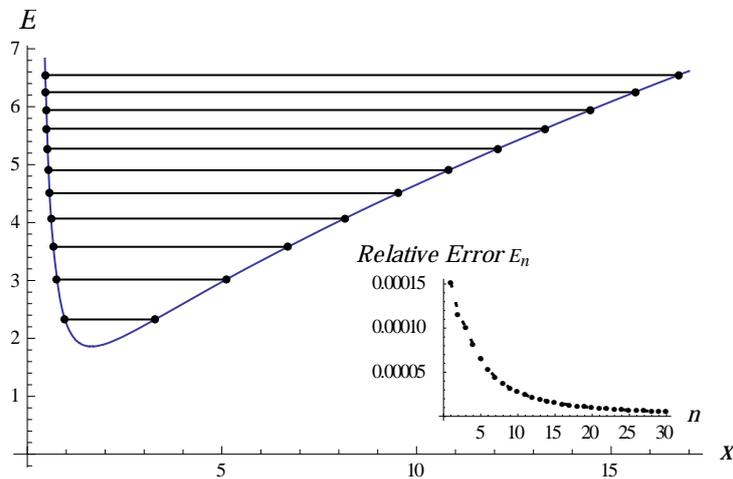

Fig. 4. Energy eigenvalues for potential (2) ($V_1 = 1, V_0 = 0$, $m = \hbar = 1$).

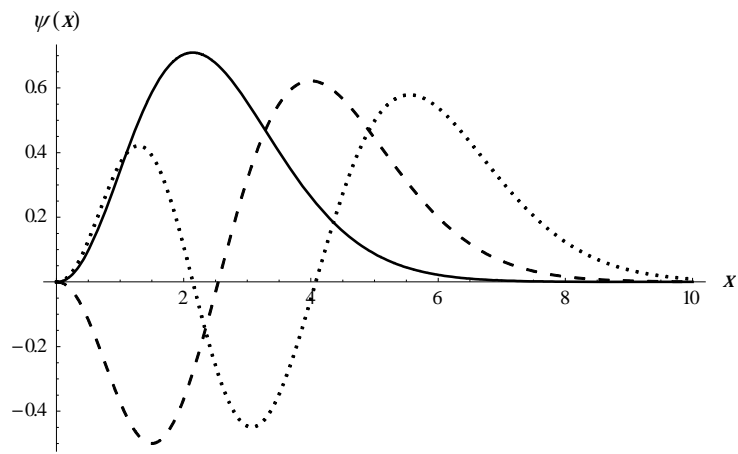

Fig. 5. Normalized wave functions for the first three bound states ($V_1 = 1$, $V_0 = 0$, $m = \hbar = 1$).

## 4. An application to nonlinear dynamics

As it was mentioned above, the potential we introduced can be applied in atomic, molecular and optical physics to describe the excitation of a two-state quantum system by a corresponding laser-field configuration. Here is an example from the theory of photo-association of ultracold atoms in Bose-condensates [40].

The semi-classical time-dependent two-state problem governing the one-color coherent photo-association of atoms is written as a system of first-order nonlinear differential equations for probability amplitudes $a_1(t)$ and $a_2(t)$ of atomic and molecular states [40]:

$$i\frac{da_1}{dt} = U(t)e^{-i\delta(t)}a_1^* a_2,$$
$$i\frac{da_2}{dt} = \frac{U(t)}{2}e^{i\delta(t)}a_1 a_1,$$
(15)

where the amplitude-modulation function $U(t)$ is the Rabi frequency and the derivative $\delta_t$ of the phase modulation function $\delta(t)$ stands for the field frequency detuning from the atom-molecule transition frequency. It is readily checked that the number of atoms plus twice the number of molecules is preserved: $|a_1|^2 + 2|a_2|^2 = \text{const} = I_N$; we put $I_N = 1$. The system (15) is equivalent to the second-order ordinary differential equation

$$\frac{d^2 a_2}{dt^2} + \left(-i\delta_t - \frac{U_t}{U}\right)\frac{da_2}{dt} + U^2 a_2 = 2U^2 |a_2|^2 a_2.$$
(16)

It has been previously shown that a highly accurate approximation for the molecular state probability $p(t) = |a_2|^2$ can be constructed via the following two-term variational ansatz [41]:

$$p = p_0(A^*, t) + C^* \frac{p_L(U_0^*, \delta_0^*, t)}{p_L(U_0^*, \delta_0^*, \infty)},$$
(17)

where $p_0$ is the solution of the polynomial equation

$$\frac{U_0^2}{\delta_z^{*2}(z)} = \frac{p_0(p_0 - \beta_1)(p_0 - \beta_2)}{9(p_0 - \alpha_1)^2(p_0 - \alpha_2)^2},$$
(18)

where $z = z(t) = \int (U(t')/U_0)dt'$, and $p_L(U_0^*, \delta_0^*, t)$ is the solution of the corresponding linear two-state problem. The ansatz involves four variational constants: $A^*, C^*, U_0^*, \delta_0^*$, the other parameters $\alpha_{1,2}$ and $\beta_{1,2}$ being determined through these parameters. Since $p_0(A^*, z)$ is a root of the quartic equation (18) and thus is known for any excitation scheme, the problem is thus reduced to the solution of the linear counterpart of equations (15). This solution is derived using the above solution of the Schrödinger equation as follows. Consider the linearization of equation (16) achieved by neglecting the nonlinear term on its right-hand side. For a constant $U = U_0$, applying the transformation

$$a_2 = e^{i\delta/2} \psi(z) \tag{19}$$

reduces the linear problem, rewritten for the variable $z$, to the following normal form with missed first-derivative term:

$$\frac{d^2\psi}{dz^2} + \left( U_0^2 + \frac{\delta_z^2}{4} + \frac{i\delta_{zz}}{2} \right)\psi = 0. \tag{20}$$

This equation is a Schrödinger equation with $U_0^2$ acting as the energy and the function

$$V(z) = -(\delta_z^2/4) - i(\delta_{zz}/2) \tag{21}$$

being the potential. For the potential (2) we discuss, resolving the last equation with respect to $\delta_z$, returning back to the time variable, and further passing to a finite duration amplitude modulation, we arrive at a field configuration shown in Fig.6. As it is seen, this is a model of pulsed excitation with an asymmetric crossing of the resonance. The time variation of the Rabi frequency is controlled by its maximal value $U_0$ and the variation of the detuning depends on a single parameter $\delta_0$. Apart from the variation limits $\delta_t(\mp\infty)$, the latter parameter also controls the asymmetry of the resonance crossing. The absence of a separate parameter controlling the asymmetry is a result of just two variable parameters (i.e., $V_0$ and $V_1$) involved in potential (2).

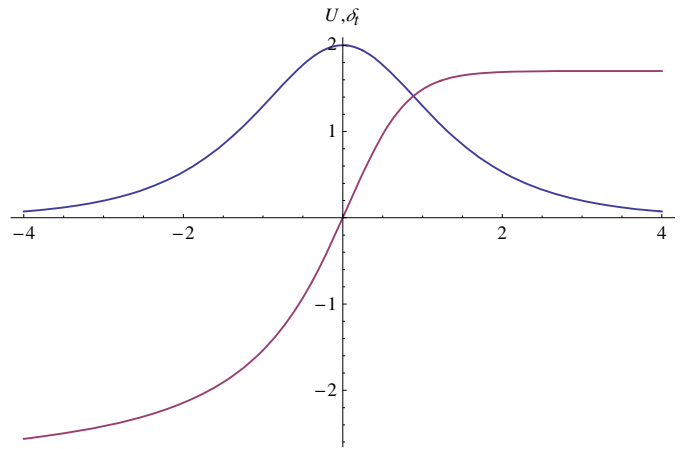

Fig. 6. Driving optical field configuration.

Discussing the conversion to the molecular state, an interesting result comes out if the final transition probability at $t \to +\infty$ is examined for $\delta_0 \gg 1$. The result is shown in Fig.7. It is seen that the transition probability reveals nonlinear saturation. This result is analytically verified by using the exact nonlinear integral equation obeyed by the probability

$$p - 4p^2 + 4p^3 = \frac{U^2}{4}\left\{\left(\int_{-\infty}^{t}\cos(\delta)(1-8p+12p^2)dx\right)^2 + \left(\int_{-\infty}^{t}\sin(\delta)(1-8p+12p^2)dx\right)^2\right\}, \quad (22)$$

which is conveniently rewritten as an integro-differential equation:

$$p(1-2p)^2 = \frac{1}{U^2}\left(\frac{dp}{dt}\right)^2 - \left(\int \delta_t dp\right)^2. \quad (23)$$

Using the linear counterpart of this equation written for an effective Rabi frequency $W$ [42]:

$$2p_L(1-2p_L) = \frac{1}{W^2}\left(\frac{dp_L}{dt}\right)^2 - \left(\int \delta_t dp_L\right)^2,$$

one is led to the following cubic polynomial equation for the final transition probability

$$p_\infty^2 - 2p_\infty^3 \approx \frac{a_0}{U_0^2}\left(p_\infty - \frac{p_{L\infty}}{2}\right) \quad (24)$$

where $a_0$ is a fitting constant. The appropriate root of this equation is presented in Fig.7 by filled circles. For large field intensities, a simple approximation for this root reads

$$p_\infty \approx \frac{1}{2} - \frac{2-p_{L\infty}}{4}\sqrt{\frac{2a}{U_0^2}}. \quad (25)$$

This is an advanced approximation for $\lambda = U_0^2 \gg 1$. Numerical testing shows that it describes the asymptotic transition probability at $t \to +\infty$ with accuracy of the order of $10^{-3}$.

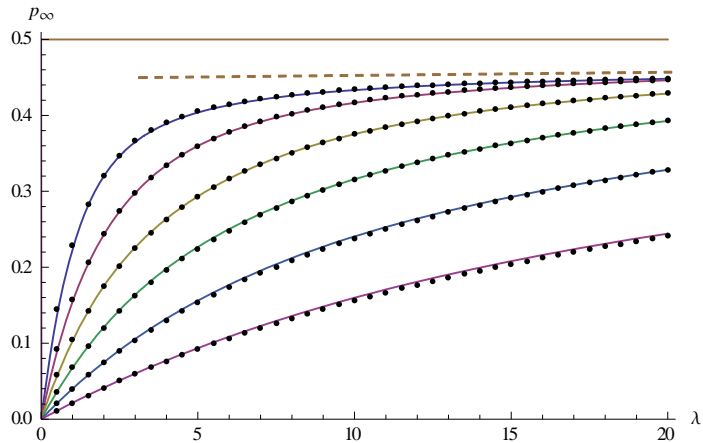

Fig. 7. Atom-molecule conversion versus $\lambda = U_0^2$ at $\delta_0 = 3, 6, 10, 20, 40$ (from the left to the right). Filled circles – numerical result, solid line – a root of Eq. (24).

## 5. Discussion

Thus, we have presented a new conditionally integrable potential for which the Schrödinger equation is solved in terms of the Hermite functions. These functions belong to the confluent hypergeometric class. The potential presents an infinitely extended confining well formed by an attractive $V_1 x^{2/3}$ term with arbitrary $V_1$ and a fixed-strength repulsive centrifugal-barrier term $\sim x^{-2}$. The general solution of the problem is composed using two fundamental solutions each of which presents an irreducible linear combination of two non-integer-order Hermite functions. The coefficients of the combinations are not constants. We have derived the exact equation for the bound-state energy spectrum and have constructed a highly accurate approximation for the energy levels. The accuracy of the approximation prevails that of the semiclassical result by two orders of magnitude. A peculiarity of the wave-functions describing the bound-states is that they are not quasi-polynomials.

The potential belongs to a remarkable six-parameter bi-confluent Heun family derived by Lemieux and Bose [30]. In its general form the family involves four power-law terms proportional to $x^{2/3}, x^{-2/3}, x^{-4/3}, x^{-2}$. As already mentioned above, the particular cases of this potential have been studied by several authors on diverse occasions. The first appearance of a confluent hypergeometric sub-potential of this family is the first Stillinger potential, which in the units $m = \hbar = 1$ is written as $V_{S1} = V_1 x^{2/3} + V_2 x^{-2/3} - 5 x^{-2}/72$ [15]. This potential, for which the fundamental solution is written using a single Hermite function, was afterwards discussed by de Souza Dutra, Znojil, Grosche and others in the context of the concept of conditional solvability (see, e.g., [43-45]). It has been shown that the energy spectrum derived through polynomial reductions do not provide physically permissible wave functions because these solutions do not vanish in the origin [44,45].

Another example of a confluent hypergeometric sub-potential of the mentioned bi-confluent Lemieux-Bose family worth being mentioned here is the second Exton potential $V_{E2} = V_1 x^{2/3} + 9 V_3^2 x^{-2/3}/2 + V_3 x^{-4/3} + 7 x^{-2}/72$ [33]. Also here, the author suggested an energy spectrum equation based on the polynomial reduction of his series solution of the corresponding Schrödinger equation [33]. However, it is shown, by a straightforward inspection, that the polynomial solutions here also do not produce wave-functions satisfying the correct boundary condition $\psi(0) = 0$.

Of course, the above examples by no means discredit the very idea of polynomial solutions. Indeed, the mentioned Lemieux-Bose potential as well as many other (multi) confluent or general Heun potentials do allow polynomial reductions which provide physically correct useful solutions. Such solutions have been applied, e.g., in the context of quasi-exactly solvability [46,47] or as possible interaction models in particle physics, quantum chemistry, atomic and laser physics, nanophysics, etc. [48-52]. The above examples, including the potential that we have just introduced, merely indicate the need to carefully examine the infinite series solutions as well.

**Acknowledgments**

This work has been supported by the Armenian National Science and Education Fund (Grant No. PS-4986), the Armenian Science Committee (SC Grants No. 18RF-139 and No. 18T-1C276), and the Russian-Armenian (Slavonic) University in the expense of the Ministry of Education and Science of the Russian Federation. The work by T.A. Ishkhanyan was partially supported by SPIE through a 2017 Optics and Photonics Education Scholarship. T.A. Ishkhanyan thanks the French Embassy in Armenia for a doctoral grant as well as the Agence universitaire de la Francophonie with the Armenian Science Committee for a Scientific Mobility grant.

**References**

1. E. Schrödinger, "Quantisierung als Eigenwertproblem (Erste Mitteilung)", Annalen der Physik **76**, 361-376 (1926).
2. E. Schrödinger, "Quantisierung als Eigenwertproblem (Zweite Mitteilung)", Annalen der Physik **79**, 489-527 (1926).
3. P.M. Morse, "Diatomic molecules according to the wave mechanics. II. Vibrational levels", Phys. Rev. **34**, 57-64 (1929).
4. C. Eckart, "The penetration of a potential barrier by electrons", Phys. Rev. **35**, 1303-1309 (1930).
5. G. Pöschl, E. Teller, "Bemerkungen zur Quantenmechanik des anharmonischen Oszillators", Z. Phys. **83**, 143-151 (1933).
6. N. Rosen and P.M. Morse, "On the vibrations of polyatomic molecules", Phys. Rev. **42**, 210-217 (1932).
7. M.F. Manning and N. Rosen, "A potential function for the vibrations of diatomic molecules", Phys. Rev. **44**, 953-953 (1933).
8. L. Hulthén, Ark. Mat. Astron. Fys. **28A**, 5 (1942); L. Hulthén, "Über die Eigenlösungen der Schrödinger-Gleichung der Deuterons", Ark. Mat. Astron. Fys. **29B**, 1-12 (1942).
9. R.D. Woods and D.S. Saxon, "Diffuse surface optical model for nucleon-nuclei scattering", Phys. Rev. **95**, 577-578 (1954).
10. F. Scarf, "New soluble energy band problem", Phys. Rev. **112**, 1137-1140 (1958).
11. A.M. Ishkhanyan, "The third exactly solvable hypergeometric quantum-mechanical potential", EPL **115**, 20002 (2016).
12. A.M. Ishkhanyan, "Exact solution of the Schrödinger equation for the inverse square root potential $V_0/\sqrt{x}$", EPL **112**, 10006 (2015).

13. A.M. Ishkhanyan, "The Lambert W-barrier - an exactly solvable confluent hypergeometric potential", Phys. Lett. A **380**, 640-644 (2016).
14. A.M. Ishkhanyan, "A singular Lambert-W Schrödinger potential exactly solvable in terms of the confluent hypergeometric functions", Mod. Phys. Lett. A **31**, 1650177 (2016).
15. F.H. Stillinger, "Solution of a quantum mechanical eigenvalue problem with long range potentials", J. Math. Phys. **20**, 1891-1895 (1979).
16. R. Dutt, A. Khare and Y.P. Varshni, "New class of conditionally exactly solvable potentials in quantum mechanics", J. Phys. A **28**, L107-L113 (1995).
17. J.N. Ginocchio, "A class of exactly solvable potentials. I. One-dimensional Schrödinger equation", Ann. Phys. **152**, 203-219 (1984).
18. C. Grosche, "Conditionally solvable path integral problems: II. Natanzon potentials", J. Phys. A **29**, 365-383 (1996).
19. G. Junker and P. Roy, "Conditionally exactly solvable problems and non-linear algebras", Phys. Lett. **232**, 155-161 (1997).
20. G. Lévai, B. Konya, Z. Papp, "Unified treatment of the Coulomb and harmonic oscillator potentials in D dimensions", J. Math. Phys. **39**, 5811-5823 (1998).
21. A. Ishkhanyan and V. Krainov, "Discretization of Natanzon potentials", Eur. Phys. J. Plus **131**, 342 (2016).
22. A. López-Ortega, "New conditionally exactly solvable inverse power law potentials", Phys. Scr. **90**, 085202 (2015).
23. A. López-Ortega, "A conditionally exactly solvable generalization of the potential step", arXiv:1512.04196 [math-ph] (2015).
24. A. López-Ortega, "New conditionally exactly solvable potentials of exponential type", arXiv:1602.00405 [math-ph] (2016).
25. T.A. Ishkhanyan and A.M. Ishkhanyan, "Solutions of the bi-confluent Heun equation in terms of the Hermite functions", Ann. Phys. **383**, 79-91 (2017).
26. A.M. Ishkhanyan, "Schrödinger potentials solvable in terms of the general Heun functions", Ann. Phys. **388**, 456-471 (2018).
27. M. Znojil, "Symmetrized exponential oscillator", Mod. Phys. Lett. A **31**, 1650195 (2016).
28. M. Znojil, "Morse potential, symmetric Morse potential and bracketed bound-state energies", Mod. Phys. Lett. A **31**, 1650088 (2016).
29. R. Sasaki and M. Znojil, "One-dimensional Schrödinger equation with non-analytic potential $V(x) = -g^2 \exp(-|x|)$ and its exact Bessel-function solvability", J. Phys. A **49**, 445303 (2016).
30. A. Lemieux and A.K. Bose, "Construction de potentiels pour lesquels l'équation de Schrödinger est soluble", Ann. Inst. Henri Poincaré A **10**, 259-270 (1969).
31. H. Exton, "The exact solution of two new types of Schrodinger equation", J. Phys. A **28**, 6739-6741 (1995).
32. X. Song, "An effective quark-antiquark potential for both heavy and light mesons", J. Phys. G: Nucl. Part Phys. **17**, 49-55 (1991).
33. A. Ronveaux (ed.), *Heun's Differential Equations* (Oxford University Press, London, 1995).
34. S.Yu. Slavyanov and W. Lay, *Special functions* (Oxford University Press, Oxford, 2000).
35. F.W.J. Olver, D.W. Lozier, R.F. Boisvert, and C.W. Clark (eds.), *NIST Handbook of Mathematical Functions* (Cambridge University Press, New York, 2010).
36. A. Hautot, "Sur des combinaisons linéaires d'un nombre fini de fonctions transcendantes comme solutions d'équations différentielles du second ordre " Bull. Soc. Roy. Sci. Liège **40**, 13-40 (1971).

37. G. Szegö, *Orthogonal Polynomials*, 4th ed. (Amer. Math. Soc., Providence, 1975).
38. A.M. Ishkhanyan and V.P. Krainov, "Maslov index for power-law potentials", JETP Lett. **105**, 43-46 (2017).
39. C. Quigg and J.L. Rosner, "Quantum mechanics with applications to quarkonium", Phys. Rep. **56**, 167 (1979).
40. K. Goral, M. Gadja, and K. Rzazewski, "Multimode dynamics of a coupled ultracold atomic-molecular system", Phys. Rev. Lett. **86**, 1397-1401 (2001).
41. A. Ishkhanyan, B. Joulakian, and K.-A. Suominen, "Variational ansatz for the nonlinear Landau-Zener problem for cold atom association", J. Phys. B **42**, 221002 (2009).
42. A. Ishkhanyan, "Demkov-Kunike model in cold molecule formation", Proc. SPIE **7998**, 79980T (2011).
43. A. de Souza Dutra, "Conditionally exactly soluble class of quantum potentials", Phys. Rev. A **47**, R2435-R2437 (1993).
44. M. Znojil, "Comment on "Conditionally exactly soluble class of quantum potentials", Phys. Rev. A **61**, 066101 (2000).
45. C. Grosche, "Conditionally solvable path integral problems", J. Phys. A **28**, 5889-5902 (1995).
46. A.G. Ushveridze, *Quasi-exactly solvable models in quantum mechanics* (IOP Publishing, Bristol, 1994).
47. A.V. Turbiner, "One-dimensional quasi-exactly solvable Schrödinger equations", Physics Reports **642**, 1-71 (2016).
48. S.K. Bose, "Exact bound states for the central fraction power singular potential", Nuovo Cimento **109 B**, 1217-1220 (1994).
49. O.V. Veko, K.V. Kazmerchuk, E.M. Ovsiyuk, V.V. Kisel, A.M. Ishkhanyan, V.M. Red'kov. J. Nonlinear Phenomena in Complex Systems. **18**, 243-258 (2015).
50. T.A. Shahverdyan, D.S. Mogilevtsev, A.M.Ishkhanyan, and V.M. Red'kov. J. Nonlinear Phenomena in Complex Systems. **16**, 86-92 (2013).
51. J. Karwowski and H.A. Witek, "Bi-confluent Heun equation in quantum chemistry: Harmonium and related systems", Theor. Chem. Accounts **133**, 1494 (2014).
52. F. Caruso, J. Martins, V. Oguri, "Solving a two-electron quantum dot model in terms of polynomial solutions of a bi-confluent Heun equation", Annals of Physics **347**, 130 (2014).